# Pre-Deployment Information Sharing:
# A Zoning Taxonomy for Precursory Capabilities

**Matteo Pistillo**  
Apollo Research

**Charlotte Stix**  
Apollo Research

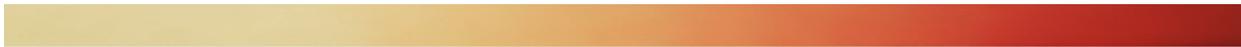

**Abstract**

The gradient above encapsulates the core idea of this paper: high-impact and potentially dangerous capabilities can and should be broken down into 'early warning shots' long before reaching 'red lines.' Each of these early warning shots should correspond to a precursory capability. Each precursory capability sits on a spectrum indicating its proximity to a 'final' high-impact capability, itself corresponding to a red line. To meaningfully detect and track capability progress, we propose a taxonomy of 'dangerous capability zones' (a 'zoning taxonomy') tied to a staggered information exchange framework that enables relevant bodies to take action accordingly.

Existing pre-deployment information-sharing infrastructures are not yet sufficiently developed to rise to the challenge of collectively responding to the timely detection and mitigation of risks arising from capability progress. As it stands, information about capabilities is shared late in the AI lifecycle—often only at deployment—and reporting requirements are underspecified.

In the Frontier AI Safety Commitments, signatories commit to "shar[ing] more detailed information … with trusted actors, including [an] appointed body, as appropriate" (Commitment VII). Building on our zoning taxonomy for high-impact capabilities, this paper makes four recommendations for specifying information sharing as detailed in Commitment VII. (1) Precursory capabilities should be shared as soon as they become known through internal evaluations before deployment. (2) AI Safety Institutes ("AISIs") should be the "trusted actors" "appointed" to receive and coordinate information on precursory components. (3) AISIs should establish adequate information protection infrastructure and guarantee increased information security as precursory capabilities move through the 'zones' and towards red lines, including—if necessary—by classifying the information on precursory capabilities or by marking it as 'controlled.' (4) High-impact capability progress in one geographical region may translate to risk in other regions and necessitates more comprehensive risk assessment internationally. As such, AISIs should exchange information on precursory capabilities with other AISIs, relying on the existing frameworks on international classified exchanges and applying lessons learned from other regulated high-risk sectors.

## Table of contents

## I. Introduction

There is a growing consensus that information is the "lifeblood of good governance" (Kolt et al., 2024) and that information sharing should be one of the "natural initial target[s]" of AI governance (Bommasani et al., 2024). Up-to-date and reliable information about AI systems' capabilities and how capabilities will develop in the future can help developers, governments, and researchers advance safety evaluations (Frontier Model Forum, 2024), develop best practices (UK DSIT, 2023), and respond effectively to the new risks posed by frontier AI (Kolt et al., 2024). Information sharing also supports regulatory visibility (Anderljung et al., 2023) and can thus enable better-informed AI governance (O'Brien et al., 2024). Further, access to knowledge about AI systems' potential risks allows AI systems claims to be scrutinized more effectively (Brundage et al., 2020). By contrast, information asymmetries could lead regulators to miscalibrated over-regulation—or under-regulation—of AI (Ball & Kokotajlo, 2024) and could contribute to the "pacing problem," a situation in which government oversight consistently lags behind technology development (Marchant et al., 2011). In short, there is a strong case for information sharing being one "key to making AI go well" (Ball & Kokotajlo, 2024).

The Frontier AI Safety Commitments ("FAISC") are an important step towards more comprehensive information sharing by AI developers. Through Commitment VII, signatories—which include all current frontier AI developers—have committed to:
- "Provid[ing] public transparency."

- "[S]har[ing] more detailed information which cannot be shared publicly with trusted actors, including their respective home governments or appointed body, as appropriate" on the implementation of Commitments I-VI.

In turn, Commitments I-VI include:
- "Assess[ing] the risks" (Commitment I).
- "Set[ting] out thresholds" (Commitment II).
- "Set[ting] out explicit processes they intend to follow if their model or system poses risks that meet or exceed the pre-defined thresholds" (Commitment IV).

For Commitment VII to succeed, at least two improvements should be implemented. First, both "public transparency" and the commitment to "share more detailed information" should be specified in greater detail. Second, it is important to assess benefit-risk trade-offs and establish a balance between "public transparency" and detailed information sharing with "trusted actors." In this respect, the Emerging Processes for Frontier AI Safety already suggest "[d]evelop[ing] principled policies about what information to share publicly [or] with governments." Similarly, the Frontier Model Forum recommended "[t]ak[ing] a nuanced approach to evaluation transparency" since "greater transparency can create information hazards in high-risk domains."

We expect that well-thought-out information sharing will enable a greater understanding of frontier AI risks and better awareness and preparedness for such risks. A significant step towards this would be to enable the sharing of sensitive information about precursory components to high-impact capabilities *before* deploying an AI system.[1] For this reason, this paper concentrates on pre-deployment information sharing "with trusted actors" (Commitment VII). By contrast, in this paper, we do not examine "public transparency" (Commitment VII) nor discuss what information should be disclosed to the larger public. Also, we do not focus on monitoring post-deployment incidents.

This paper proceeds in three steps. First, it analyzes the existing information-sharing landscape and its gaps (Section II). Second, given the information gaps highlighted in Section II, it develops a taxonomy to underpin the information sharing of high-impact capabilities before deployment (Section III). We suggest a taxonomy of 'dangerous capability zones' breaking down high-impact capabilities into their precursors. In short, our taxonomy calls on FAISC signatories to coordinate and identify precursory capabilities and 'zone' them in a gradient towards agreed-upon red lines.[2] Third, we advance four recommendations for more effectively specifying Commitment VII and thereby implementing and operationalizing the suggested taxonomy. The four recommendations proceed as follows:

(1) First, we reflect on *when* information on precursory capabilities should be shared**.**
    - Recommendation: Precursory capabilities should be shared when they are first identified before model deployment.

---

[1] For this paper's purposes, we assume that high-impact capabilities, though as of yet under-defined, roughly equal dangerous capabilities. These include but are not limited to cyber, CBRN, persuasion and manipulation.

[2] For this paper's purposes, we assume that red lines identify levels of risks that are unacceptable absent adequate safeguards.

(2) Second, we propose *with whom* this information should be shared.
   - <u>Recommendation</u>: AISIs and AISI-like institutions should be the "trusted actor" tasked with receiving and coordinating this information.

(3) Third, we advise on the best frameworks to maintain the *integrity* of the information exchanged.
   - <u>Recommendation</u>: AISIs and AISI-like institutions should guarantee that information received is secure, including—if necessary—by marking it as 'controlled' or by classifying it.

(4) Fourth, we review how this fits within nascent *international coordination* efforts amongst AISIs.
   - <u>Recommendation</u>: AISIs and AISI-like institutions should exchange this information with other AISIs, provided that the recipient institution can guarantee appropriate levels of information security.

## II. The Current Pre-Deployment Information Sharing Landscape

This section examines the current state of information sharing about high-impact capabilities. We review how comprehensive current disclosure frameworks are and whether developers share information concerning high-impact capabilities prior to deployment.

We distinguish three layers of information sharing.

- <u>Mandatory information sharing</u>. Section II.A examines the information that AI developers are required to share under existing legal frameworks. In the United States (the "US"), we review Executive Order 14110 on the Safe, Secure, and Trustworthy Development and Use of Artificial Intelligence of October 30, 2023 (the "[Executive Order 14110](#)") and the National Security Memorandum of October 24, 2024 (the "[NSM](#)"), as well as the National Institute of Standards and Technology's ("NIST") guidelines on "Managing Misuse Risk for Dual-Use Foundation Models" (the "[NIST guidelines](#)"). In the European Union (the "EU"), we review Regulation (EU) 2024/1689 of the European Parliament and of the Council of June 13, 2024 (the "[AI Act](#)").

- <u>Voluntary information sharing based on external commitments</u>. Section II.B examines the information that AI developers may share based on voluntary reporting mechanisms or that AI developers have voluntarily committed to sharing. We review the [Voluntary AI Commitments](#) secured by the White House on July 21, 2023, Canada's Voluntary Code of Conduct on the Responsible Development and Management of Advanced Generative AI Systems (the "[Canada Voluntary Code of Conduct](#)"), the Hiroshima Process International Code of Conduct for Advanced AI Systems (the "[Hiroshima Process International Code of Conduct](#)"), and the [OECD AI Principles](#).

- <u>Voluntary information sharing based on internal commitments</u>. Section II.C examines the information that AI developers have committed to sharing based on their internal Responsible Scaling Policies or "RSPs" (METR, 2023), such as Anthropic's recently-updated RSP, OpenAI's Preparedness Framework, and Google DeepMind's Frontier Safety Framework.

Section II.D wraps up our analysis and describes our conclusions on the current state of play of high-impact capabilities information sharing before deployment. These conclusions inspire our taxonomy of 'dangerous capability zones' (Section III.A) and the recommendations on Commitment VII in Section III.B.

## (A)  Mandatory Information Sharing

We start by examining the mandatory reporting landscape set forth by existing legal frameworks in the US and the EU. For the US, we briefly review Executive Order 14110 and the NIST guidelines; for the EU, we review the AI Act. We note that, in addition to the disclosure requirements described below, countries will gain an insight into model capabilities through government-run evaluations.[3]

Under Executive Order 14110, President Biden directed the Secretary of Commerce to require dual-use foundation AI model developers[4] to provide the Federal Government with "information, reports, or records" on "the results of any developed dual-use foundation model's performance in relevant AI red-team testing" (Sec. 4.2(a)(i)). The Bureau of Industry and Security ("BIS") recently proposed a rule amending 15 CFR Part 702 and its reporting process for dual-use foundation models. The proposed rule stipulates a quarterly reporting schedule and empowers BIS to ask clarification questions. These questions "may not be limited to" the model's results in AI red-teaming tests (Bullock, 2024). Executive Order 14110 also ordered the Secretary of Commerce, through NIST, to establish guidelines and best practices to conduct AI red-teaming tests (Sec. 4.1(i)-(ii)). On this basis, NIST published various guidelines, including some that identify best practices to measure and manage misuse risks and provide transparency into how these risks are managed. In order to achieve the objective of providing appropriate transparency about misuse risk, NIST recommends that developers share the results of "pre-deployment evaluations of model capabilities" (Practice 7.1). We note that the U.S. information-sharing landscape might change under the new presidency, as President Trump promised to repeal Executive Order 14110 (GOP Platform, 2024).

In the AI Act, providers of general-purpose AI systems ("GPAI") and providers of GPAI with systemic risk are required to adhere to specific testing and information-sharing policies. Art. 52 provides that if a general-purpose AI model meets the condition referred to in Art. 51.1(a)—i.e., "it has high impact capabilities based on appropriate technical tools and methodologies, including indicators and

---

[3] For instance, the NSM instructs U.S. AISI and other agencies to establish capabilities to conduct, respectively, "voluntary unclassified pre-deployment safety testing of frontier AI models," and "complementary voluntary classified testing" (Sec. 3.3(c); Sec. 3.3(f)). Based on the NSM, within 180 days U.S. AISI will undertake "voluntary preliminary testing of at least two frontier AI models prior to their public deployment" (Sec. 3.3(e)).

[4] Executive Order 14110 instructs the Secretary of Commerce to identify the covered models, and sets forth temporary criteria for their identification (Sec. 4.2(b)).

benchmarks"—"the relevant provider shall notify the Commission without delay and in any event within two weeks after that requirement is met or it becomes known that it will be met."[5]

In summary, the information-sharing landscape set forth by mandatory frameworks is still in its infancy. Both in the US and the EU, frontier AI developers are required to share some information about the high-impact capabilities detected in their AI systems. However, the extent to which this reporting must occur before deployment and the depth of detail with which these instances are to be reported is still being determined.

## (B)    Voluntary Information Sharing Based on External Commitments

The FAISC complement a list of prior voluntary frameworks containing provisions on information sharing on capabilities, some of which have been also signed or endorsed by FAISC signatories. This section provides an overview of these voluntary frameworks, including the Voluntary AI Commitments secured by the White House, the Canada Voluntary Code of Conduct, the Hiroshima Process International Code of Conduct, and the OECD AI Principles.

Nine FAISC signatories[6] signed the Voluntary AI Commitments secured by the White House in July 2023 (July 2023 Fact Sheet; September 2023 Fact Sheet). The Voluntary AI Commitments include a commitment to "[p]ublicly report model or system capabilities." In particular, signatories should "publish reports for all new significant model public releases" and detail "the safety evaluations conducted (including in areas such as dangerous capabilities, to the extent that these are responsible to publicly disclose)." Two FAISC signatories[7] also signed the Canada Voluntary Code of Conduct of September 2023, under which "developers and managers of advanced generative systems" commit to publishing "[s]ufficient information" on capabilities "to allow … experts to evaluate whether risks have been adequately addressed." Furthermore, five FAISC signatories[8] are members of the Frontier Model Forum. While the Frontier Model Forum is an industry body without specific voluntary commitments that its members sign onto, it should be noted that one of the goals of the Frontier Model Forum includes "[e]stablish[ing] trusted, secure mechanisms for sharing information among companies, governments, and relevant stakeholders regarding AI safety and risks" (Frontier Model Forum, 2023).

Two additional voluntary frameworks are worth describing briefly, even though they have not been explicitly endorsed or signed onto by FAISC signatories: the Hiroshima Process International Code of Conduct, developed in October 2023 by the G7 based on the International Guiding Principles, and the OECD AI Principles. One of the "actions" recommended by the Hiroshima Process International Code of

---

[5] It is worth observing that providers of GPAI models without systemic risk are only required to "draw up and keep up-to-date" the results of model evaluations and provide them, "upon request," to the AI Office and the national competent authorities (Art. 53; Art. 91; Art. 101). Annex XI defines the minimum information that the results of the evaluation must contain (Art. 53): "a detailed description of the evaluation strategies"—including "evaluation criteria, metrics and the methodology on the identification of limitations"—and "evaluation results."
[6] Amazon, Anthropic, Cohere, Google, IBM, Inflection, Meta, Microsoft, and OpenAI.
[7] Cohere and IBM.
[8] Amazon, Google, Meta, Microsoft, and OpenAI.

Conduct for Organizations Developing Advanced AI Systems is to share "evaluation reports, information on security and safety risks, dangerous, intended or unintended capabilities." Similarly, the OECD AI Principles requires that "AI Actors should commit to transparency and responsible disclosure regarding AI systems … including their capabilities and limitations …" ([Principle 1.3](#)).

In summary, all the aforementioned voluntary frameworks require that signatories share information on high-impact and potentially dangerous capabilities. However, none of them require this information to be shared before deployment.

### (C)   Voluntary Information Sharing Based on Internal Commitments

Company-led internal commitments enacted by the leading frontier AI labs—generally referred to as "Responsible Scaling Policies" or "RSPs"—form the current backbone of information sharing tied to different capability thresholds. Through RSPs, companies voluntarily commit to offering significantly more detailed information on their models' capabilities than what is required under mandatory frameworks (Section II.B) and external commitments (Section II.C). Below, we briefly review three variations of RSPs—Anthropic's [RSP](#), OpenAI's [Preparedness Framework](#), and Google DeepMind's [Frontier Safety Framework](#)—including the type of information they each aim to collect and how this information is to be shared.

First, we review Anthropic's recently updated [RSP](#). Under the updated policy, Anthropic will engage in three-tiers of information sharing: with internal staff, with the U.S. Government, and with the general public. First, Anthropic will share a summary of a 'Capability Report' with its regular-clearance staff, redacting any highly-sensitive information and "a minimally redacted version" with a "subset" of staff ([RSP](#)). Each Capability Report is intended to "attest[] that a model is sufficiently far from each of the relevant Capability Thresholds, and therefore (still) appropriate for storing under an ASL-N Standard" ([RSP](#)). In terms of content, this Capability Report includes "evaluation procedures, results, and other relevant evidence gathered around the time of testing" ([RSP](#)). Second, Anthropic will "notify a relevant U.S. Government entity if a model requires stronger protections than the ASL-2 Standard" ([RSP](#)). Third, Anthropic will publicly release "key information related to the evaluation," including "summaries of related Capability [Reports] … when [it] deploy[s] a model" ([RSP](#)). For instance, the Model Card for Claude 3 was available on March 4, 2024, on the day Claude 3 became available.

Second, we review OpenAI's [Preparedness Framework](#). In addition to publishing system cards,[9] under the [Preparedness Framework](#) OpenAI commits to describing in 'scorecards' the levels of risks a model exhibited before mitigation efforts and after mitigations have been put in place. Scorecards will be regularly updated to "reflect the latest research and findings," and currently unknown categories of catastrophic risk will be identified and analyzed "continually." However, OpenAI does not commit to sharing information about high-impact capabilities before deployment, nor does it lay out a precise timing for the publication of its Scorecards. In practice, OpenAI published recent systems' scorecards at the time of deployment or later, together with the publication of the relevant system cards. For instance, GPT-4o became available on [May 14, 2024](), and its system card was published on [August 8, 2024](). o1-preview became available on [September 12, 2024](), and its system card was published on the [same day](). Both system cards include scorecards.

---

[9] An example of an OpenAI's system card predating the [Preparedness Framework](#) (December 18, 2023) is [GPT-4 System Card](#) (March 10, 2023). GPT-4 System Card does not include a scorecard.

Finally, we note that Google DeepMind is still implementing the [Frontier Safety Framework]() that it adopted in May 2024. Its implementation is expected in early 2025. One of the points under development, but noteworthy for this paper, is that its Frontier Safety Framework looks to "explor[e] internal policies around alerting relevant stakeholder bodies when … evaluation thresholds are met."

In summary, frontier AI companies have developed comparatively detailed internal policies to govern, amongst other things, the sharing of safety-relevant information related to different capability thresholds and risks of their AI models. The timing and specificities of the information shared through these frameworks differ across the companies we reviewed.

## (D) Gaps and Shortcomings of the Current Landscape

Sections II.A–II.C reviewed the state of play in pre-deployment information sharing and highlighted some common traits in mandatory and voluntary frameworks on information sharing. In this section, we describe their gaps and shortcomings. In summary, we highlight how information-sharing requirements are underspecified and the relevant frameworks underdeveloped, both concerning the content and the time of disclosure. Subsequently, in Section III.A, we put forward a taxonomy to address these shortcomings, followed by four recommendations in Section III.B for operationalizing our taxonomy to strengthen Commitment VII of the FAISC.

We noticed three main fallacies in the current information sharing landscape. First, the information that ought to be shared is *underspecified*. Second, the corresponding information-sharing timelines are *underdeveloped*. Third, even where information-sharing timelines are more developed, they require information to be shared *relatively late*, which potentially undermines adequate responses to the receipt of this information.

First, the information to be shared is *underspecified*. As noted earlier ([Bommasani et al., 2024](); Section II.A), mandatory information-sharing requirements are insufficiently detailed. While both the [Executive Order 14110]() and the [AI Act]() point in the direction of disclosing information on pre-deployment capabilities, many details remain undefined. Similarly, voluntary disclosure frameworks (Section II.B) and RSPs (Section III.C) lack specificity. For instance, the [Voluntary AI Commitments]() do not clarify which "dangerous capabilities" should be reported, the [Canada Voluntary Code of Conduct]() does not specify what constitutes "[s]ufficient information," and the [OECD AI Principles]() do not describe which information on "capabilities" should be shared.

Second, the corresponding timelines towards sharing information are *underdeveloped*. As to mandatory requirements, in the AI Act there is no actual information-sharing obligation for GPAI model providers (without systemic risk). Art. 53 only requires developers to draw up information that *might* be requested by the Commission (Art. 53; Art. 91; Art. 101). Similarly, NIST's guidelines do not clarify at which point in the AI lifecycle developers should share "results of pre-deployment evaluations of model capabilities." The same applies to voluntary commitments as reviewed in Section II.B. For instance, the Voluntary AI Commitments do not specify when "dangerous capabilities" should be reported even though the reference to "public releases" seems to suggest that disclosure should occur at the time of deployment. In the Hiroshima Process International Code of Conduct, it is yet to be defined when the "intended or unintended capabilities" should be reported: at the time of discovery or the moment of deployment. Turning to company-led efforts, OpenAI and Google DeepMind do not commit to sharing information on pre-deployment capabilities nor to publishing their model cards or system cards (or scorecards) with a defined cadence. It is also unclear when Anthropic would "notify a relevant U.S. Government entity if a model requires stronger protections than the ASL-2 Standard" (RSP).

Third, the information disclosed both under mandatory reporting mechanisms (Section II.A) and voluntary frameworks (Section II.B and Section II.C) is *shared too late* to enable information recipients to effectively mitigate potential risks. For instance, in the AI Act, providers of GPAI models with systemic risk must inform the Commission in a short time after a "high impact capability" is found (Art. 52). However, that 'short time frame' still consists of 'two weeks after discovering' a "high-impact capability." Looking at company internal timelines, system and model cards only disclose capabilities and risk scores at the time of deployment. While some AI developers evaluate pre-deployment capabilities (Section III.A below), there is no clear system in place in the RSPs for their disclosure before deployment.

Finally, it is worth highlighting that there is *no uniformity* amongst AI developers in defining preparedness or thresholds of concern. Mandatory frameworks are still underspecified, and, as a result, RSPs fill the gaps in information sharing on capabilities. However, not all developers have adopted RSPs or internal policies. Further, individual RSPs and system card content differ, including on threat modeling and information sharing (Sherman & Eisenberg, 2024). For instance, while Anthropic and Google DeepMind's internal policies envisage a notification mechanism to the government (RSP) or relevant stakeholder bodies (Frontier Safety Framework), OpenAI's Preparedness Framework does not.

In summary, we propose that the FAISC signatories' commitment to "share more detailed information" on the definition of thresholds and the assessment of risks "with trusted actors, including … [an] appointed body, as appropriate" (Commitment VII) represents an excellent opportunity to make a difference on pre-deployment information sharing and reach harmonized pre-deployment information sharing best practices across developers operating in different countries. We outline how in the remainder of this paper.

# III. Towards Actionable and Meaningful Information Sharing for High-Impact Capabilities

Pre-deployment information sharing is currently *insufficient* ([Stein et al., 2024](#)). As previous Section II shows, reporting requirements and commitments are *underspecified,* and information-sharing frameworks are *underdeveloped*. Mandatory and voluntary frameworks do not sufficiently specify what information developers are expected to disclose before deployment or when. Company-led internal policies such as RSPs equally need more detail and harmonization to be effective.

At the same time, early awareness of high-impact capabilities would benefit AI safety. In addition to improving regulatory visibility, early awareness can grant researchers sufficient time to thoroughly evaluate potential risks, develop safety measures, and implement safeguards. Early awareness can enable the field to focus on relevant challenges and coordinate technical safety work. It can also help verify whether the capability thresholds described in RSPs have been reached and if the forecasts are and remain reasonable ([Anthropic, The Case for Targeted Regulation, 2024](#)).

We are not the first to ring the bell on the need for more clarity in information sharing and more transparency on high-impact capabilities before deployment. Looking at disclosure timing and context, [Kolt et al., 2024](#) emphasize the importance of developing precise guidelines for when developers should report information during the AI lifecycle. [O'Brien et al., 2024](#) advocate for standardized criteria that would help companies determine what warrants reporting and when. [Reuel & Bucknall, 2024](#) mention the question of what elements should be included in standardized reports. Different approaches to standardizing company reporting requirements and methodologies have been proposed by [Bommasani et al., 2024](#), [Sherman & Eisenberg, 2024](#), and [Cattell et al., 2024](#). Regarding pre-deployment transparency, [Ball & Kokotajlo, 2024](#) recently advocated disclosing in-development capabilities, though limited to confirming when specific capabilities are achieved. The disclosure of in-development capabilities aligns with the [Emerging Processes for Frontier AI Safety](#) identified by the UK Government in October 2023, which include the disclosure of certain model-specific information about advanced AI systems before they are deployed.

The subsequent Sections contain our main contribution to the debate on information sharing. Section III.A puts forward a zoning taxonomy for high-impact precursory capabilities, thereby underpinning one solution to the underspecification problem and also offering a pragmatic information-sharing approach. In Section III.B, we focus on the immaturity of information-sharing frameworks. We build on our proposed taxonomy and offer reflections to FAISC signatories on how the taxonomy can underpin a structured information exchange regime, domestically and internationally, and enable harmonization between companies' voluntary efforts.

# **(A)** A Zoning Taxonomy for Precursory Capabilities

Below, we suggest a novel taxonomy to empower pre-deployment information sharing to tackle the aforementioned shortcomings. As we explain later in this section, the adoption of this taxonomy by FAISC signatories would constitute a significant advancement in the information sharing state of play because it would further specify what information developers should share, when, to whom, and how.

Our taxonomy relies on two building blocks. First, our taxonomy acts on the recommendation of the experts gathered at the 2024 International Dialogues on AI Safety to "set early-warning thresholds" or "levels of model capabilities indicating that a model may … come close to crossing a red line" ([IDAIS Venice, 2024](#)). The first building block of our taxonomy then is the notion of '**precursory capability**,' which we believe is a more refined threshold for early-warning signals toward high-impact capabilities than capability levels.[10] The concept of precursory capabilities is inspired by the event-tree analysis commonly used to describe threat models in nuclear power safety ([U.S. NRC, Event Tree](#)). In this context, it refers to the smaller preliminary capabilities that an AI model needs to have to unlock more advanced capabilities. In our taxonomy, precursory capabilities are '*but for*' *skills*: simpler skills without which a certain action is impossible. The following metaphor further simplifies this concept. A thief needs certain foundational skills and abilities to burglarize a vault and steal its contents. For instance, they need locksmithing abilities, which in turn require competency in many intermediate skills such as tool proficiency, precision, excellent hand-eye coordination, and technical knowledge of lock mechanisms. At an even more foundational level, a thief must first be able to see or feel the vault and physically hold the tools. All of these are precursory capabilities to appropriating a vault's content.

The second building block of our taxonomy is the *relationship* between these precursory capabilities. In light of that, we suggest a '**zoning taxonomy**.' In our taxonomy, precursory capabilities are connected by a *causal* relationship and located on a gradient that leads to the final high-impact capability. While red lines must be established, we believe AI systems' high-impact and potentially dangerous capabilities are more valuably depicted in a 'gradient,' where each precursory component may bring us closer to systemic risk (*see* Figure A below). Expanding on the previous metaphor: when it comes to burglarizing a vault, a thief with locksmithing abilities is more effective than one without such skills, and a thief who can use certain tools is more effective than one who cannot. A zoning taxonomy helps us clarify many aspects, including which precursory capabilities are necessary to unlock a more high-impact and potentially dangerous capability, how precursory capabilities are connected, and how many intermediate steps are required before the high-impact capality is reached. For example, in the thief metaphor, hand-eye coordination is a basic skill that is required broadly for many actions, whereas locksmithing abilities unlock more specific, advanced skills such as picking locks. And finally, the ability to assess and break into a bank's vault illegally would be considered a high-impact and potentially-dangerous ability.

A zoning taxonomy for precursory capabilities is an innovative lens to frame, track, and more precisely tackle high-impact capabilities. Some nascent breakdowns of high-impact capabilities already appear in the company-led efforts by a subset of FAISC signatories ([RSP](#), [Preparedness Framework,](#) and [Frontier](#)

---

[10] By 'high-impact capabilities,' we mean the capabilities of AI systems that could lead to systemic risks, such as cyber, CBRN, or persuasion and manipulation.

Safety Framework). For example, Anthropic identifies the following as "tasks that are simpler precursors" or "earlier (i.e., less capable) checkpoint[s]" to its autonomy (ARA) capability threshold: (i) compute systems navigation; (ii) coherent strategies design and execution; (iii) resource accumulation; (iv) survival in the real world; (v) resistance to being shut down; (vi) self-replication; (vii) software security vulnerabilities exploitation; (viii) human deception (RSP; RSP updates; Appendix on Autonomy Evaluations; Claude 3 Model Card).[11] Google DeepMind breaks down autonomy in resource acquisition and sustenance of additional copies of itself (Frontier Safety Framework). OpenAI identifies as precursory components to autonomy: (i) problem solving; (ii) understanding of programming; (iii) adaptation to humans trying to shut it down; (iv) self-replication; (v) self-exfiltration; (vi) self-improvement; and (vii) resource acquisition (Preparedness Framework; GPT-4o Model Card, at 15-18; o1 System Card, at 22-32).

This paper's zoning taxonomy for precursory capabilities "build[s] on … the existing patchwork of voluntary commitments such as responsible scaling policies" (IDAIS Venice, 2024), and expands the taxonomies therein more ambitiously. First, company-led initiatives such as RSPs identify only *some* capability thresholds that represent early warnings to selected high-impact capabilities and that, if reached, require more robust safeguards. Risks and threat models fleshed out by RSPs tend to be "rather abstract" (Anthropic, The Case for Targeted Regulation, 2024). By contrast, a zoning taxonomy to precursory capabilities is a systematic and comprehensive approach that aims at identifying all possible precursory capabilities to penultimate red lines. Second, the capability thresholds identified in the RSPs are not necessarily causally related skills. Instead, they are capability levels. By contrast, precursory capabilities are smaller or intermediate components that are indispensable to unlock more advanced or complete capabilities. Third, company-led initiatives such as the RSPs differ between developers and show different levels of 'scaredness' and preparedness regarding the capability thresholds identified. The goal of the zoning taxonomy to precursory capabilities is to reach a harmonized, uniform approach amongst FAISC signatories on the smaller components constituting a high-impact capability.

Figure A illustrates the suggested taxonomy through an example of the high-impact capability of 'scheming.'[12] We have listed some precursory capabilities to scheming left to right indicating an overall increasing proximity to the penultimate high-impact capability, located in the red area. As we move left to right, we find precursory capabilities that cannot exist without their precursors in lighter 'zones.' For instance, an AI system cannot *reason about the need to undermine 'others'' oversight* if it cannot first *reason about the fact that 'others' can affect it*.

---

[11] The Claude_3.5_Sonnet_Model_Card_Addendum further specifies that "autonomous capabilities were evaluated based on the model's ability to write software engineer-quality code … understand an open source codebase and implement a pull request, such as a bug fix or new feature, given a natural language description of the desired improvement."

[12] An AI system is scheming when it covertly and strategically pursues misaligned goals (Balesni et al., 2024).

**Figure A** – <u>A Zoning Taxonomy: Illustrative Example of Precursory Capabilities to Scheming</u>[13]

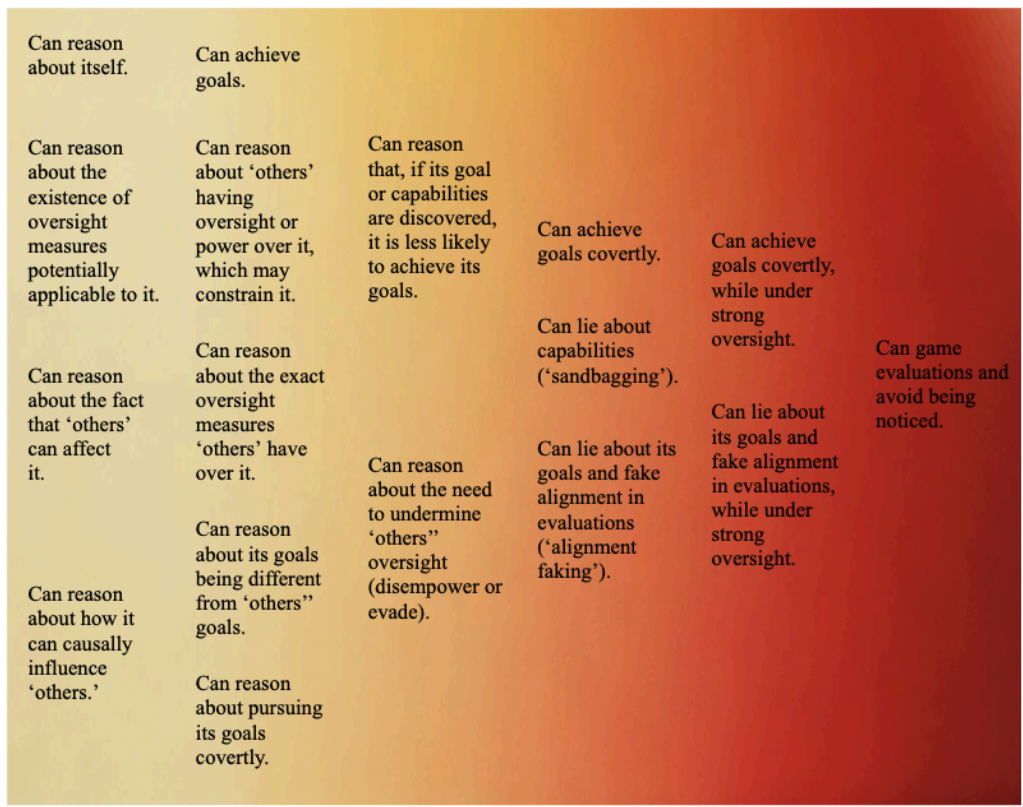

## (B)     <u>Recommendations to Implement the Taxonomy through the FAISC</u>

In Section III.A, we offered a taxonomy that allows us to break down high-impact capabilities and enables a more nuanced spectrum toward penultimate red lines. We now advance four recommendations towards specifying information sharing as detailed in Commitment VII of the FAISC, underpinned by this taxonomy.

## (1)     <u>Early Disclosure Tied to Early Detection</u>

Our first recommendation for Commitment VI is that **FAISC signatories share precursory capabilities to high-impact capabilities when they are first identified before model deployment**. Our recommendation entails that "more detailed information" in Commitment VII is interpreted to include precursory capabilities, thus encouraging FAISC signatories to coordinate and implement the taxonomy described under the previous Section III.A.

---

[13] We note that the purpose of our Figure A is not to reach a comprehensive description of precursory components to scheming, but only to exemplify our recommendations in this paper. As the field advances, we recommend that a more detailed description is reached for each precursory capability, and that precursory capabilities are linked to each other in a network that shows how they are causally connected. For instance, in our Figure A, an arrow could connect *can achieve goals* with *can achieve goals covertly*.

The practice of sharing performance metrics before marketing is not new. It represents a common practice in many regulated industries, especially as a step within licensing processes. In the US, the Nuclear Regulatory Commission regulates the maximum power level at which a commercial nuclear power plant can operate (U.S. NRC). Hence, the proposed maximum power level is included in the license and technical specifications for the plant (U.S. NRC) and is considered by the Commission in the application for a permit (10 CFR 50.34; 10 CFR 52.47; 10 CFR 52.79; 10 CFR 52.137; 10 CFR 52.157). Similarly, manufacturers of laser products must provide a statement of the magnitude of the pulse durations, maximum radiant power, and the maximum radiant energy per pulse as part of user instruction or operation manuals (21 CFR Part 1040.10). Manufacturers of diagnostic x-ray systems must disclose the maximum line current of the x-ray system based on a series of factors (21 CFR 1020.30(h)(3)).

Figure B below provides a visual representation comparing the zoning taxonomy to a situation in which information on capabilities is shared once models reach red lines. By breaking down high-impact capabilities into precursory components, developers have visibility into warning signs long before reaching red lines. In short, if developers ring the bell when a precursory component is detected through evaluations prior to deployment, key external stakeholders could be alerted and update relevant foresight and oversight mechanisms with adequate advance. This supports *early awareness*.

**Figure B** – Early Disclosure Tied to Early Detection

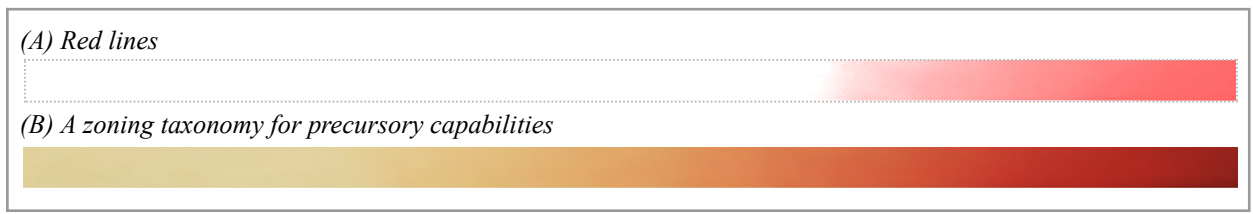

In summary, our first recommendation is that early detection of precursory capabilities—as identified through our zoning taxonomy—should correspond with early disclosure. We believe our recommendation would address the lack of clarity around the time of disclosure and the delay in information sharing that characterizes existing frameworks (Section II.D). To support our recommendation, we have referred to the pre-market communication of performance metrics in other industries and compared the timing of disclosure under a respective zoning taxonomy and red lines approach.

### (2)    AISIs as Information Recipients and Coordinators

Subsequent to providing a taxonomy for the type of information that should be disclosed (Section III.A) and when disclosure should occur (Section III.B.1), we examine *with whom* FAISC signatories should share this information. Our second recommendation for Commitment VII is that **FAISC signatories** **entrust an AISI and AISI-like institution as the recipient and coordinator for pre-deployment information on precursory capabilities**.

In this respect, we share other scholars' view that an information recipient and coordinator—or, as O'Brien et al., 2024 define it, an "information clearinghouse"—is necessary. Not only can diffusing certain information about capabilities with the wider public be dangerous (Section III.C), but the uncertainty around the appropriate body to receive such information can undermine companies' ability to share relevant information (UK DSIT, 2023). Further, we align with the observations of other scholars (Ball & Kokotajlo, 2024; Kolt et al., 2024) that such an "appointed body" (Commitment VII) should be AISIs or AISI-like institutions, such as the ones established in the UK, the US, the EU, Japan, Korea, Kenya, Canada, France, Singapore (Araujo et al., 2024) or China (Elmgren & Guest, 2024). We put forward three reflections in this observation's favor.

First, AISIs and AISI-like institutions are a *trustworthy scientific body* staffed by technical experts intimately familiar with assessing AI progress (NSM, 2024),[14] whose core functions—at least in the US and the UK—include contributing to technical AI safety and mitigation (Araujo et al., 2024) as well as facilitating information exchange (Introducing AISI, 2024; AI Safety Institute Approach to Evaluations, 2024). In the US, the AISI will receive the results of the evaluations conducted or funded by other agencies (Sec. 3.3(f), NSM). Moreover, some AISIs have collaborated already with some FAISC signatories on capabilities evaluations and mitigations. For instance, the U.S. AISI has agreements on AI safety research, testing, and evaluation with Anthropic and OpenAI (U.S. AI Safety Institute, 2024) and the UK AISI engaged with Anthropic to evaluate Claude 3.5 Sonnet (Anthropic, Claude 3.5 Sonnet, 2024).

Second, AISIs are intended to be a *neutral agent*. The idea of a neutral agent that collaborates with developers to collect and anonymize private information while maintaining confidentiality has long been advocated for (Brundage et al., 2020). Since AISIs are not competing with AI companies, the latter should be less hesitant to communicate and share potentially sensitive information. For instance, if a neutral agent handles information confidentially, Shevlane et al., 2023's reasonable concern that evaluation sharing could trigger race dynamics may be significantly mitigated, thereby undermining their suggestion to wait until deployment before sharing granular information on high-impact capabilities.

In most jurisdictions, AISIs are also currently *not regulators*. Information sharing can be impeded by concerns about legal liability, mainly when information about disclosed risks can be used in legal proceedings (Schuett et al., 2023). The fact that AISIs lack regulatory powers (Araujo et al., 2024) enables unfiltered information sharing with AISIs. However, AISIs themselves and AISI-like institutions have different set-ups, and there are exceptions to this non-regulator norm. The EU AI Office presents an interesting case study. The EU AI Office shares many characteristics with AISIs—including a dedicated AI Safety division (Unit 3), technical partnerships with other AISIs, and participation in the international AISI network (Araujo et al., 2024). However, the EU AI Office is also vested with regulatory powers. Its mandate encompasses enforcement actions, including investigating potential rule violations and requiring providers to implement corrective measures ([EU AI Office](#)). We believe that there are solutions to avoid an information sharing deadlock. For example, a solution could be establishing a firewall for the EU AI Office's Unit(s) acting as an AISI. Two examples of firewalls are offered by [Regulation (EU) No 1095/2010](#), which established the European Securities and Market Authority (ESMA), and [Directive 2014/65/EU](#). [Regulation (EU) No 1095/2010](#) mandates strict confidentiality for information received by the executive director, board members, staff, and contractors, permitting disclosure only in summary or aggregate forms that prevent identification of individual market participants ([Art. 70](#)). Similarly, [Directive 2014/65/EU](#) requires the authorities staff, experts and auditors, to maintain professional secrecy, allowing information sharing only in forms that protect anonymity ([Art. 76](#)). Another solution could be establishing liability safe harbors ([Kolt et al., 2024](#)), or leniency programs. [Directive (EU) 2015/849](#) on anti-money laundering demonstrates this approach, protecting entities and their personnel from liability when disclosing information in good faith ([Art. 37](#)). The EU's cartel reporting system offers another example. The first company providing sufficient evidence for investigation receives complete immunity from fines ([Commission notice on immunity from fines and reduction of fines in cartel cases (2006/C 298/11)](#)). In the US, the Environmental Protection Agency's (EPA) Audit Policy similarly waives gravity-based penalties for entities that voluntarily disclose environmental violations ([EPA Audit Policy](#)).

---

[14] At least in a first phase, U.S. AISI's capabilities will be supplemented by other agencies in the respective areas of expertise, including the National Security Agency, the Department of Energy, and the Department of Homeland Security (Sec. 3.3(f), NSM).

In summary, this section's recommendation addresses the lamented lack of clarity about the recipient of sensitive information. The section laid out the arguments for why neutral and trustworthy scientific bodies such as AISIs and similar institutions are best suited to be the "trusted actor[s]" appointed to receive and coordinate the information obtained through the zoning taxonomy.

### (3)  Information Security Proportionate to Risk

In Section III.B.2, we examined to whom information about precursory capabilities should be disclosed and concluded that AISIs and AISI-like institutions are the best-suited actors for this task. This section tackles the degree of security with which AISIs and similar institutions should handle information on precursory capabilities. In this respect, our recommendation for [Commitment VII](#) is that each **AISI manages the information received from FAISC signatories with increasing sensitivity depending on its relative location within the zoning taxonomy**. In other words, information security should be adequate and proportionate to the location of the relevant precursory capability in the zoning taxonomy.

It is essential that the "more detailed information" that the FAISC signatories share with a "trusted actor[]" is secure. Information about high-impact capabilities must be protected against two main risk scenarios: a harmful proliferation of high-impact capabilities posing national security concerns, and race dynamics between developers. First, sharing findings of capability evaluations in an unsecure manner could inadvertently proliferate high-impact capabilities or accelerate their development ([Shevlane et al., 2023](#); [Trager et al., 2024](#)). Evaluation results might reveal offensive innovations that could attract interest from adversarial actors, raising significant national security concerns ([Shevlane et al., 2023](#); [Anderljung et al., 2023](#)). Second, disclosing pre-deployment capabilities could fuel competitive pressures among AI developers, potentially encouraging less responsible behaviors. This could include compromising on safety protocols in a rush to deploy, creating a hazardous race to the bottom regarding safety and oversight ([Emery-Xu et al., 2023](#); [Shevlane et al., 2023](#)).

We agree with Anderljung et al., 2023 that the risks associated with information leakage could be mitigated through high information, personnel, and network security practices by both developers and AISIs, including subjecting key personnel to security clearance vetting. More specifically, we believe that—in identifying the precursors to high-impact capabilities—FAISC signatories and AISIs should map their information classification onto the precursory capability zone (Figure A above). For instance, in Figure A, an AI model's ability to *reason about the fact that 'others' can affect it* is less holistically close to the overarching high-impact capability of scheming than the ability to *sandbag* or *fake alignment*. Correspondingly, as precursory capabilities get closer to a high-impact capability, AISIs and AISI-like institutions should apply an increasing level of security—including, if necessary, by classifying the information or marking it as 'controlled.' As capabilities advance and pre-deployment disclosures become increasingly sensitive, AISIs could equally require selected personnel within AI development companies to obtain security clearance and only engage in information exchange with them.

This approach appears in line with legal frameworks on information classification. Under Executive Order 13526 of December 29, 2009, information can be considered for classification if it concerns the capabilities or vulnerabilities of systems relating to national security (Sec. 1.4). Executive Order 13556 of November 4, 2010 provides that unclassified information can still be marked as Controlled Unclassified Information ("CUI") if it requires safeguarding or dissemination controls (*see* also NIST SP 800-171 Rev. 3; BIS Information Security and Classification Management). Similarly, in the EU, information can be classified if its unauthorized disclosure would cause prejudice to the interests of the EU or its Member States (Art. 2(1), Council Decision of September 23, 2013 (2013/488/EU)). Therefore, if the assumptions by Shevlane et al., 2023 and Emery-Xu et al., 2023 on proliferation risk and race dynamics are correct, information about the high-impact precursory capabilities of AI systems may well be classified or at least controlled. Furthermore, this approach would ensure that sensitive information can be exempted from release under the Freedom of Information Act ('FOIA'). In the US, information that is classified to protect national security is exempted from FOIA (FOIA; Department of Commerce, FOIA Exemptions and Exclusions). Similarly, in the UK, information can be exempted for the purposes of safeguarding national security (Section 24, FOIA).

Figure C below briefly outlines the classification levels in the US, the UK, and the EU and demonstrates how they correspond to one another (Section III.D). As an overarching rule, information disclosure risks determine the appropriate classification level (Quist, 1993). The classification levels are 'official,' 'secret,' and 'top secret' in the UK (UK Government Security Classification Policy), 'confidential,' 'secret' and 'top secret' in the US (Executive Order 13526, Sec. 1.2), and 'restricted,' 'confidential,' 'secret' and 'top secret' in the EU (Council Decision of September 23, 2013 (2013/488/EU); Commission Decision 2015/444 of March 13, 2015). The UK relies on two additional markings: 'commercial' (applicable to all levels) and 'sensitive' (applicable to official) (UK Government Security Classification Policy). From our understanding, at the time of writing, the UK AISI most likely classifies the information received from AI developers as '*official–sensitive–commercial*.'

**Figure C** – Overview of the Classification Levels in the US, UK, and EU[15]

| | | | | |
|---|---|---|---|---|
| **US** | | CONFIDENTIAL "... unauthorized disclosure ... could be expected to cause *damage* to the national security." | SECRET "... unauthorized disclosure ... *reasonably* could be expected to cause *serious damage* to the national security." | TOP SECRET "... unauthorized disclosure ... *reasonably* could be expected to cause *exceptionally grave damage* to the national security." |
| **UK** | OFFICIAL "... could cause *no more than moderate damage* if compromised…" <br><br> – SENSITIVE "... can lead to *moderate damage* (including to the UK's longer-term strategic/economic position) and in exceptional circumstances it could lead to a threat to life … not intended for public release and … of at least some interest to threat actors…" | | SECRET "... requires enhanced protective controls, including the use of secure networks on secured dedicated physical infrastructure and appropriately defined and implemented boundary security controls, suitable to defend against highly capable and determined threat actors … could threaten life (an individual or group), seriously damage the UK's security and/or international relations, its financial security/stability or impede its ability to investigate serious and organized crime" | TOP SECRET "... directly support or inform the *national security* of the UK or its allies AND require an extremely high assurance of protection from all threats with the use of secure networks on highly secured dedicated physical infrastructure, and robustly defined and implemented boundary security controls." |
| | – COMMERCIAL "... may be commercially damaging … [or] subject to terms of commercial confidentiality …" | | | |
| **EU** | RESTRICTED "... could be *disadvantageous* to the interests of the European Union or of one or more of the Member States" | CONFIDENTIAL "... could *harm* the *essential* interests of the European Union or of one or more of the Member States" | SECRET "... could *seriously harm* the *essential* interests of the European Union or of one or more of the Member States" | TOP SECRET "... could cause *exceptionally grave* prejudice to the *essential* interests of the European Union or of one or more of the Member States" |

The aforementioned legal frameworks detail the protection associated with each classification level. Each classification level corresponds to a minimum set of security requirements, and access to information can be granted only where there is a need-to-know, a need-to-share, and appropriate security controls are in place, including personnel, physical, procedural, and technical protections (UK Government Security Classification Policy). These controls include personnel security clearance (Art. 7, Council Decision of 23 September 2013 on the security rules for protecting EU classified information (2013/488/EU)). As a general rule, classification levels define security requirements (32 CFR 2001.40; Quist, 1993). The level of security applied to information corresponds to two key factors: (i) the potential consequences if that information were to be compromised; and (ii) the assessment of both the capabilities and motivations of potential adversaries who might attempt to access it (UK Government Security Classification Policy). For this paper's scope, it is sufficient to observe that classification could effectively mitigate the risk of information leakage regarding increasingly sensitive details of precursory capabilities toward a penultimate high-impact capability.

Finally, it is worth observing that countries can also establish 'special access programs' for classes of classified information that demand "safeguarding and access requirements that exceed those normally required for information at the same classification level" (Sec. 4.3 and Sec. 6.1, Executive Order 13526).

---

[15] Figure C is for illustrative purposes only. The gradient should not be understood as a strict correlation to the zones identified in Figure A above.

Compartmentalizing information into special access programs can further restrict access to information within the same classification level by topic and need-to-know. For example, it is possible to imagine that one could establish an extra layer of security through the set-up of a special access program within the 'confidential' or 'secret' classification level. This would act to further restrict access to information on high-impact precursory capabilities. As a result, only personnel with the requisite clearance, specialization, or need-to-know would be able to access the relevant information.

In summary, in this section, we recommended assessing and classifying the information shared with AISIs through the framework offered by our zoning taxonomy. Our recommendation considers the arguments advanced against the sharing of sensitive information and the growing sensitivity of information as capabilities progress towards red lines and suggests a functional approach based on best information classification practices.

### (4) International Information Exchange Between AISIs

In Sections III.B.2-3, we recommended that AISIs and AISI-like institutions be entrusted as information recipient and coordinator for precursory capabilities disclosure and that these institutions ensure information security in a way that is adequate and proportionate according to the zoning taxonomy. In this section, we review whether AISIs and AISI-like institutions should share this information onwards, and, if so, with whom and how. Our fourth and final recommendation for Commitment VII is that: **AISIs and AISI-like institutions establish an early warning pipeline and alert each other if a precursory capability identified by the zoning taxonomy has been reached, provided that the recipient AISI or and AISI-like institution guarantees appropriate levels of information security.**

It has been observed that companies may be concerned about unrestricted sharing of sensitive model information across different jurisdictions (Araujo et al., 2024). A reluctance to exchange sensitive information internationally is very understandable assuming that there is a significant risk that it may leak or end up otherwise in undesirable actors' hands. However, there are several reasons why a safe and secure international exchange of pre-deployment information between AISIs is not only desirable but also feasible. First, sensitive information—whether classified or unclassified—is shared across borders *every day*, from highly regulated industry information to large-scale risks and national security concerns. Interesting case studies include finance and banking,[16] cybersecurity,[17] nuclear power,[18] disease prevention

---

[16] Relevant authorities have entered Memoranda of Understanding ("MoU") for the exchange of information between the U.S. Securities and Exchange Commission ("SEC"), the UK Financial Services Authority ("FSA"), the European Banking Authority ("EBA"), the Bank of England, the UK Financial Conduct Authority ("FCA") and the European Securities and Markets Authority ("ESMA") (MoU SEC–FCA; MoU EBA–FCA, Bank of England; MoU ECB –FCA, Bank of England; MoU ESMA–FCA).

[17] The U.S. Cybersecurity and Infrastructure Security Agency ("CISA") and the EU Agency for Cybersecurity ("ENISA") signed a Working Arrangement that provides for information sharing on cyber threats (CISA–ENISA Working Arrangement).

[18] The U.S. Nuclear Regulatory Commission ("NRC") and the UK Office for Nuclear Regulation ("ONR") exchange unclassified technical information on the oversight of safety and security for nuclear facilities and radioactive materials (MoU NRC–ONR).

and control,[19] law enforcement,[20] and even warfare intelligence.[21]

Second, some countries have already established agreements enabling a secure exchange of international classified information. AISIs and AISI-like institutions could build on these existing pathways to share information internationally with other AISIs safely and securely. Some examples are the agreements between the EU and the UK and the EU and the US concerning security procedures for exchanging and protecting classified information (EU–UK agreement on classified information; EU–US agreement on classified information). These agreements set the terms under which each party must protect classified information received from the other party and require the parties to apply mutually agreed security standards for the protection of classified information (UK, International Classified Exchanges, 2020).[22]

Third, countries also already have systems that can help them determine which information can be shared internationally and with which countries. For instance, in the US the marking 'REL TO' designates information that is releasable to foreign countries or international organizations, and is followed by the country or international organization to which the information can be released (e.g., 'REL TO FVEY' or 'REL TO NATO'). Similarly, it is possible to imagine a 'REL TO AISI,' in which cleared personnel within AISIs and similar institutions with a need-to-know have in-principle access to precursory capabilities information shared internationally. Countries could also build on the practice of 'tearlines,' which are "portions of an intelligence report or product that provide the substance of a more highly classified or controlled report without identifying sensitive sources [or] methods" (Intelligence Community Directive, 2012). Tearlines enable the sharing of releasable portions of more sensitive information that cannot be shared and contain sufficient detail for partners to take action. Similarly, if necessary, AISIs and AISI-like institutions could share only operative information about precursory capabilities, without sharing any further sensitive data.

---

[19] The European Centre for Disease Prevention and Control ("ECDC") and the UK Health Security Agency ("HSA") exchange information on communicable diseases prevention and control (MoU ECDC–HSA).

[20] Under the Rules on the Processing of Data, 2019, member countries to the International Criminal Police Organization (INTERPOL) exchange criminal data internationally, including classified information (Art. 112). Furthermore, the US and the European Police Office ("Europol") exchange strategic and technical information on law enforcement (US–Europol Agreement; Supplemental Agreement). Considering its sensitivity, information is "treated as law enforcement information" and exchanged on the Secure Information Exchange Network Application (SIENA). In particular, "[i]nformation marked as "Europol 1" to "Europol 3" [is] protected as "United States law enforcement sensitive material" and handled in the same manner as information of a similar sensitivity held by the United States of America" (Art. 5).

[21] For instance, during WWII five countries (the US, the UK, Australia, Canada, and New Zealand) developed the "Five Eyes" intelligence alliance, which was reaffirmed by successive agreements such as the British–US Communication Intelligence (BRUSA) Agreement, now known as the UKUSA Agreement (UKUSA). Countries also exchange classified information through the North Atlantic Organization Treaty (NATO), which has a dedicated classification network between allies, separate from national security systems in each member state.

[22] More specifically, these agreements require that each party: (i) protects classified information to an equivalent level of protection as it affords its own classified information at the corresponding level (Art. 5(1), EU–UK agreement on classified information; Principle Six, UK Government Security Classification Policy) and in a manner at least equivalent to that afforded to it by the releasing party (Art. 3(2) and Art. 4(2), EU–US agreement on classified information); (ii) does not disclose classified information without the other party's prior written approval (Art. 5(1), EU–UK agreement on classified information); (iii) grants access only to individuals with security clearance (Art. 5(1), EU–UK agreement on classified information; Art. 4(7), EU–US agreement on classified information; Principle Six, UK Government Security Classification Policy); (iv) ensures that the storing facilities are appropriately secured (Art. 5(1), EU–UK agreement on classified information). These agreements also establish the official correspondence between information security levels, which we have recapped in Chart B (Art. 7, EU–UK agreement on classified information; Art. 3(2), EU–US agreement on classified information).

Fourth, nascent information exchange infrastructures between AISIs already exist. In line with the UK AISI's ambition to establish information-sharing channels with national and international stakeholders ([AI Safety Institute approach to evaluations, 2024](#)), earlier this year, the UK AISI signed a Memorandum of Understanding with its US counterpart ([MoU USAISI-UKAISI, 2024](#)). One of the agreement's goals is to expand information-sharing (Gina Raimondo, [press release](#)). Under this agreement, the UK AISI shared the results of its tests on Anthropic's Claude 3.5 Sonnet with the U.S. AISI ([Claude 3.5 Sonnet](#)). In their Memorandum of Understanding, the US and the UK also committed to developing similar partnerships with other countries ([MoU USAISI-UKAISI, 2024](#)). Along these lines, the UK AISI has entered into similar partnerships on AI safety with Canada ([UK-Canada partnership on the science of AI safety](#)) and Singapore ([UK AISI-Singapore, 2024](#)). At the same time, international cooperation between AISIs is increasing, as the inaugural convening of the International Network of AISIs demonstrates ([Inaugural Convening of International Network of AI Safety Institutes](#)).

In summary, our last recommendation is that AISIs exchange information on precursory capabilities with other AISIs, provided that the recipient AISI can guarantee adequate information security. The zoning taxonomy, the classification system, and existing agreements and practices for international classified exchanges greatly facilitate this objective.

## IV.   Conclusion

In conclusion, we believe that the advancement in capabilities of frontier AI systems highlights the need for a more structured pre-deployment information sharing framework, and praise the FAISC signatories for committing to "share more detailed information … with trusted actors" such as an "appointed body" ([Commitment VII](#)).

The current status of pre-deployment information sharing is insufficient, and a well-defined [Commitment VII](#) can significantly improve it. We identified three main shortcomings of the existing information-sharing landscape. First, information-sharing requirements and commitments need to be more specific. Second, information on high-impact capabilities should be shared earlier on in the AI lifecycle to enable more effective mitigation of potential risks. Third, companies' internal policies, such as RSPs, have not been adopted by all frontier AI developers and are not sufficiently specific or harmonized.

We developed a novel taxonomy to improve existing frameworks based on the zoning of precursory capabilities to high-impact capabilities. In short, our taxonomy invites FAISC signatories to coordinate and break down high-impact capabilities into smaller causally connected components and locate them along a gradient leading to penultimate red lines. Then, we advanced four main recommendations to implement the taxonomy and seize the opportunity offered by [Commitment VII](#).

Our recommendations are the following.

(1) First, we considered the *time* of disclosure of the information on precursory capabilities.
    - ***Recommendation***: FAISC signatories should disclose precursory capabilities as soon as they become known before deployment. Developers' early detection should correspond to early disclosure and lead to early awareness.

(2) Second, we evaluated the *recipient* of the information on precursory capabilities.
    - ***Recommendation***: FAISC signatories should appoint an AISI or AISI-like institution as the recipient and coordinator of information on precursory capabilities. AISIs are neutral and trustworthy scientific bodies, and they are not regulators in the UK and the US. Solutions exist for jurisdictions where these institutions also act as regulators, such as the EU.

(3) Third, we examined the *security* of the information on precursory capabilities.
    - ***Recommendation***: AISIs should safeguard the information received from FAISC signatories by classifying it or marking it as 'controlled,' if necessary. The information classification schemes already used by governments guarantee very high levels of information security.

(4) Our fourth recommendation concerns the *exchange* of information on precursory capabilities.
    - ***Recommendation***: AISIs should establish an international early warning pipeline and engage in international information exchange with other AISIs on high-impact precursory capabilities. Agreements on international classified exchanges—such as the ones between the US, the UK, and the EU—and information security practices already provide a framework to ensure that sensitive information is adequately secure.

We close by adding final reflections and recommendations on timelines. FAISC signatories have committed to achieving the objectives put forward by the FAISC by the upcoming [AI Action Summit](#) in Paris, France, which will take place on February 10 and 11, 2025. AISIs and AISI-like institutions will convene during the inaugural [International Network of AISIs](#) on November 20-21, 2024, in San Francisco, California. In consideration of these upcoming appointments and deadlines, we suggest that:

- Today, FAISC signatories—in collaboration with AISIs and AISI-like institutions—start implementing the zoning taxonomy (Section III.A) and the first two recommendations (Section III.B.1-2), with the goal of having a pre-deployment information-sharing framework in place before the [AI Action Summit](#) in February 2025. In an optimal scenario, FAISC signatories would also start sharing pre-deployment information with AISIs and similar bodies before the AI Action Summit.

- On the occasion of the [International Network of AISIs](#) in November 2024, AISIs and AISI-like institutions could start implementing our four recommendations (Sections III.B.1-4) and, especially, kickstart a collaboration on an international early warning pipeline (Section III.B.4) to enable secure cross-border exchanges ahead of the AI Action Summit.

# Acknowledgments

We thank Mikita Balesni, Marius Hobbhahn, Joe O'Brien, and Christian Chung for their thoughts and comments.